\documentclass[prx,aps,amssymb,twocolumn]{revtex4-1}
\usepackage{amsmath}
\usepackage{amssymb}
\usepackage{amsthm}
\usepackage{amsfonts}
\usepackage{listings}
\usepackage{enumerate}
\usepackage{latexsym}
\usepackage{color}
\usepackage{bm}
\usepackage{hyperref}
\hypersetup{
 pdfnewwindow=true, colorlinks=true,
 linkcolor=blue, anchorcolor=blue,
 citecolor=blue, filecolor=blue,
 menucolor=blue, urlcolor=blue}

\usepackage{psfrag}

\usepackage{bm}
\usepackage{graphicx}
\usepackage{subfigure}

\newcommand{\webirpw}{\href{https://github.com/zjwang11/IR2PW}{\texttt{IR2PW}}}
\newcommand{\webposabr}{\href{https://github.com/zjwang11/UnconvMat/blob/master/src_pos2aBR.tar.gz}{\texttt{POS2ABR }}}

\RequirePackage[normalem]{ulem} %DIF PREAMBLE
\RequirePackage{color}\definecolor{RED}{rgb}{1,0,0}\definecolor{BLUE}{rgb}{0,0,1} %DIF PREAMBLE
\input{epsf}

\begin{document}

\tolerance 10000

\newcommand{\vk}{{\bf k}}

\draft

\title{Superconductivity in unconventional metals}

%in English titles articles and words like to, on, at etc are always spelled with small letters
%\author{Goodguys}
%\affiliation{Beijing National Laboratory for Condensed Matter Physics,
%and Institute of Physics, Chinese Academy of Sciences, Beijing 100190, China}
%\affiliation{University of Chinese Academy of Sciences, Beijing 100049, China}

\author{Zhilong Yang}
%\email{zhlyang@iphy.ac.cn}
\thanks{These authors contributed equally to this work.}
\affiliation{Beijing National Laboratory for Condensed Matter Physics,
and Institute of Physics, Chinese Academy of Sciences, Beijing 100190, China}
\affiliation{University of Chinese Academy of Sciences, Beijing 100049, China}

\author{Haohao Sheng}
\thanks{These authors contributed equally to this work.}
\affiliation{Beijing National Laboratory for Condensed Matter Physics,
and Institute of Physics, Chinese Academy of Sciences, Beijing 100190, China}
\affiliation{University of Chinese Academy of Sciences, Beijing 100049, China}

\author{Zhaopeng Guo}
%\thanks{These authors contributed equally to this work.}
\affiliation{Beijing National Laboratory for Condensed Matter Physics,
and Institute of Physics, Chinese Academy of Sciences, Beijing 100190, China}
\affiliation{University of Chinese Academy of Sciences, Beijing 100049, China}

\author{Ruihan Zhang}
\affiliation{Beijing National Laboratory for Condensed Matter Physics,
and Institute of Physics, Chinese Academy of Sciences, Beijing 100190, China}
\affiliation{University of Chinese Academy of Sciences, Beijing 100049, China}

\author{Quansheng Wu}
\affiliation{Beijing National Laboratory for Condensed Matter Physics,
and Institute of Physics, Chinese Academy of Sciences, Beijing 100190, China}
\affiliation{University of Chinese Academy of Sciences, Beijing 100049, China}
\author{Hongming Weng}
\affiliation{Beijing National Laboratory for Condensed Matter Physics,
and Institute of Physics, Chinese Academy of Sciences, Beijing 100190, China}
\affiliation{University of Chinese Academy of Sciences, Beijing 100049, China}

\author{Zhong Fang}
\affiliation{Beijing National Laboratory for Condensed Matter Physics,
and Institute of Physics, Chinese Academy of Sciences, Beijing 100190, China}
\affiliation{University of Chinese Academy of Sciences, Beijing 100049, China}

\author{Zhijun Wang}
\email{wzj@iphy.ac.cn}
\affiliation{Beijing National Laboratory for Condensed Matter Physics,
and Institute of Physics, Chinese Academy of Sciences, Beijing 100190, China}
\affiliation{University of Chinese Academy of Sciences, Beijing 100049, China}

\begin{abstract}
Based on first-principles calculations, we demonstrate that 1H/2H-phase transition metal dichalcogenides $MX_2~(M={\rm Nb,Ta}; X={\rm S,Se,Te})$ are unconventional metals, which have an empty-site band of $A_1'@1e$ elementary band representation at the Fermi level. The computed phonon dispersions indicate the stability of the system at high temperatures, while the presence of the soft phonon mode suggests a phase transition to the charge density wave state at low temperatures. Based on the Bardeen-Cooper-Schrieffer theory and computed electron-phonon coupling, our calculations show that the superconductivity (SC) in NbSe$_2$ is mainly attributed to the soft phonon mode due to the half filling of the empty-site band. Accordingly, the SC has been predicted in unconventional metals TaNS monolayer and 2H-TaN$_2$ bulk with computed $T_C=$ 10 K and  26 K respectively. These results demonstrate that the unconventional metals with partial filling of the empty-site band offer an attractive platform to search for superconductors.
\end{abstract}

\maketitle

\noindent{\bf INTRODUCTION} \\

In the past decades, topological materials have attracted a lot of attention due to their novel properties in  condensed matter physics~\cite{Qi2010The,Hasan2010Topological,Bernevig2006Quantum,Zhang2009Topological,RevModPhys.90.015001,Wan2011,Weng2016Topological,Wang2013,PhysRevLett.117.236401,topotase3,sixfoldfermi,qian2019topological}. Most recently, a new kind of unconventional materials has been proposed to be topologically trivial with wannierizable valence Bloch states, while they have a set of bands from an elementary band representation (EBR) on an empty site~\cite{aBR2022,aBR2021,realspaceinvariants2021}. The unconventional insulators are also known as obstructed atomic insulators ~\cite{tqc2017,PhysRevLett.121.126402,song2020rsi,xu2021filling}. Besides, there are also many unconventional metals, such as electrides~\cite{aBR2021}, catalysis~\cite{li2022obstructed}, solid-state hydrogen storage~\cite{aBR2022}, and superconductivity, etc~\cite{lanio,ferroelecNature}. 

Transition metal dichalcogenides (TMD), such as hexagonal bilayer stackings of dichalcogenides, named 2H-$MX_2$, have received attention with the discovery of charge density wave   (CDW) and  superconductivity (SC). The CDW transition temperature decreases from around 120 K in 2H-TaSe$_2$, to 80 K in 2H-TaS$_2$, down to 30 K in 2H-NbSe$_2$ and finally no CDW in 2H-NbS$_2$~\cite{PhysRevLett.86.4382,PhysRevLett.95.117006}. The superconducting critical temperature $T_C$ increases from around 0.2 K in 2H-TaSe$_2$ up to 7.2 K and 6 K in 2H-NbSe$_2$ and 2H-NbS$_2$, respectively. Although these properties have been widely studied in literature, the origin of the SC has not been revealed yet.

\begin{figure}[!tb]
\centering
\includegraphics[width=8.5 cm]{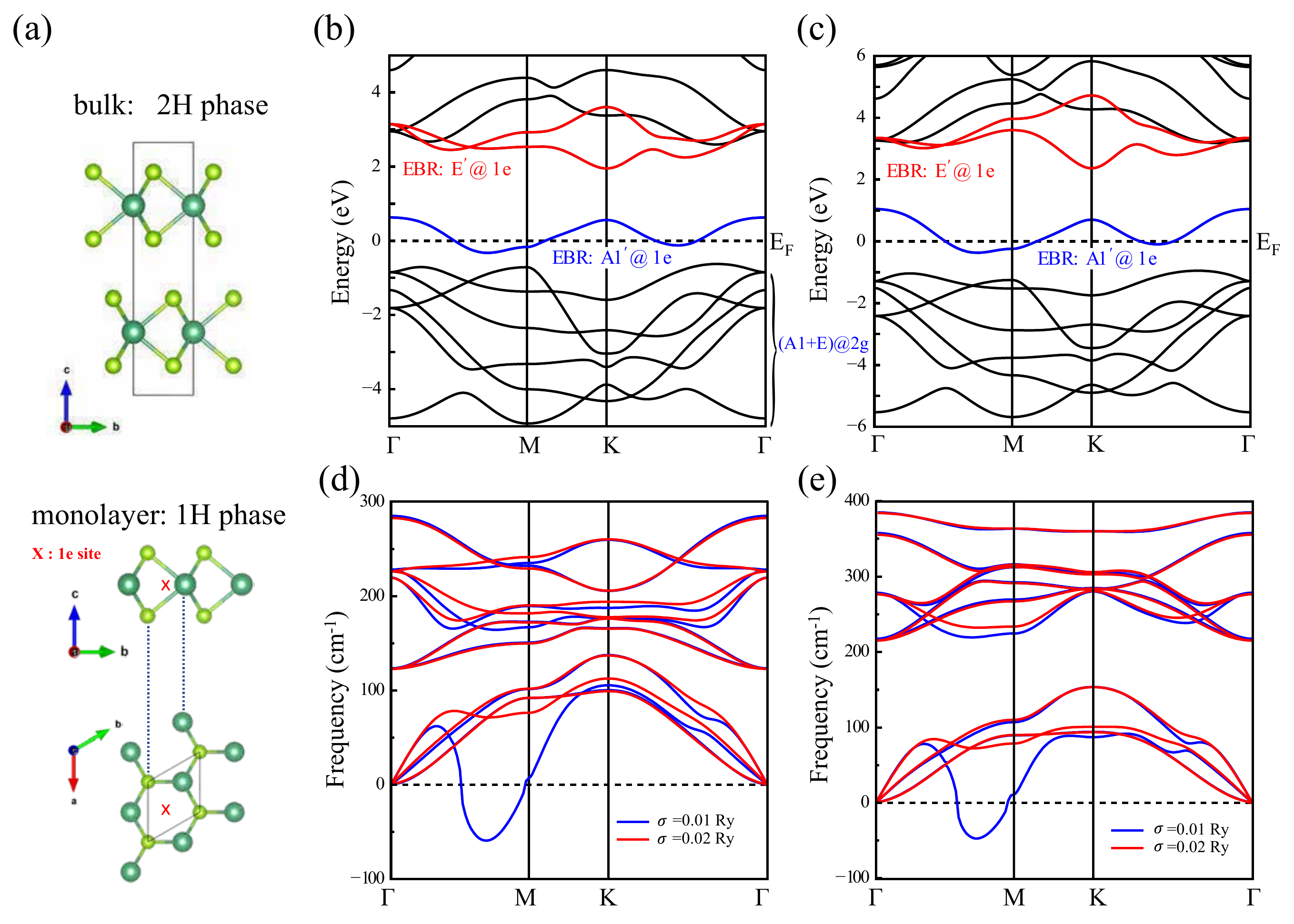}
\caption{
The crystal structures, electronic structures, and phonon dispersions of 1H-$MX_2$.
(a) Crystal structure of 1H-$MX_2$.
(b,~c) Electronic structures of NbSe$_2$ and TaS$_2$.
(d,~e) Phonon dispersions of NbSe$_2$ and TaS$_2$ with different electronic smearing parameters ($\sigma$).
} \label{fig:1}
\end{figure}

In this work, based on first-principles calculations and band representation analysis, we demonstrate that the monolayer 1H-$MX_2$ is an unconventional metal, with a half-filling EBR at an empty $1e$ site. The real space invariant (RSI) of the empty site is $\delta_1@1e=1$. The computed phonon dispersions show CDW instability at low temperatures. Based on the Bardeen-Cooper-Schrieffer (BCS) theory, the computed electron-phonon coupling (EPC) suggests the SC is attributed to soft phonon mode. The results reveal that the partial filling of the empty-site band gives rise to the strong EPC and potential SC. Following the strategy, two superconductors, TaNS monolayer and 2H-TaN$_2$ bulk, have been predicted with $T_C=$ 10 K and 26 K respectively.   \\

\begin{figure*}[!tb]
\centering
\includegraphics[width=17.5 cm]{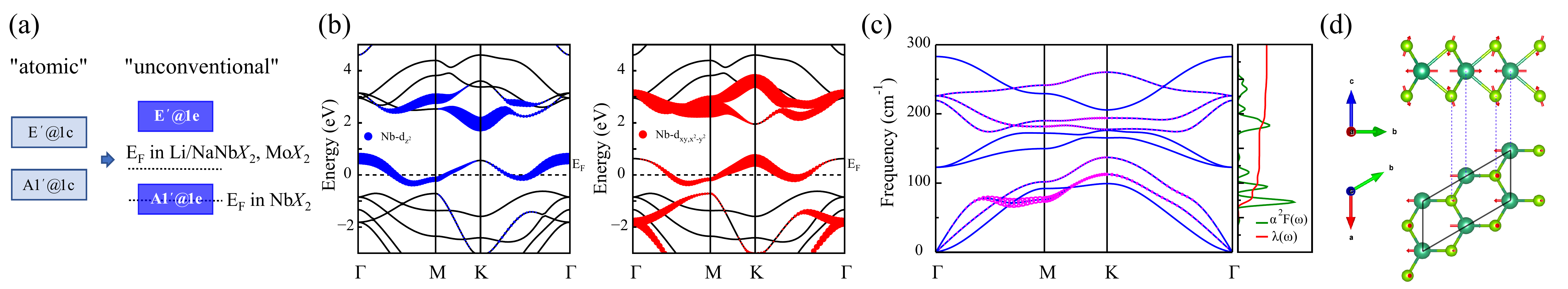}
\caption{The schematic of band hybridization, orbital-resolved band structures, and the phonon dispersions and vibration modes. (a) The schematic of the formation of the empty-site EBR $A'_1@1e$ at $E_F$. It is formed by the hybridization of $A'_1 @ 1c$ (from Nb-$d_{z^2}$) and $E' @ 1c$ (from Nb-$d_{xy,x^2-y^2}$) ABRs.
(b) Orbital-resolved band structures in 1H-NbSe$_2$ with the corresponding Nb $d_{z^2}$ and $d_{xy,x^2-y^2}$ orbitals weights colored in blue and red, respectively. The size of the symbol represents the orbital weight. (c) The phonon spectrum of 1H-NbSe$_2$ with magenta circles representing EPC $\lambda_{\bold{q}\nu}$, and the Eliashberg spectral function $\alpha_2 F(\omega)$, the frequency-dependent coupling $\lambda(\omega)$. (d) The phonon vibration mode of the lowest phonon band at $M$ point.
} \label{fig:2}
\end{figure*}

\begin{table}[b!]
    \centering
    \caption{The atomic valence-electron band representations (ABRs) of 1H-$MX_2$. The ABRs are defined as the band representations induced by the atomic valence electrons. The Nb $d$ orbitals form the $A_1^{'},E^{'}$, and $E^{''}$ irreps at the Nb($1c$) site, being $A_1^{'}@1c,E^{'}@1c$, and $E^{''}@1c$ ABRs, while the Se $p$ orbitals form the $A_1$, and $E$ irreps at the Se($2g$) site, being $A_1@2g$ and $E@2g$ ABRs.}
    \label{table:ABRs}
    \begin{ruledtabular}
    \begin{tabular}{ccccrlcc}
    %\hline\hline
        Atom & WKS($q$) & Symm. & Conf. & Irreps($\rho$) & & ABRs  &Occ. \\
        &  &  &  &  & & ($\rho@q$)  &  \\
        \hline
        Nb  & $1c$ & -62m & $d^5$ & $d_{z^2}$:&$A'_1$ & $A'_1$@1c & \\
        $(M)$ &$(\frac{1}{3}\frac{2}{3}0)$ & & & $d_{xy,x^2-y^2}$:&$E'$ &$E'$@1c & \\
        & & & & $d_{xz,yz},$:&$E''$ & $E''$@1c & \\
        \hline
        Se & $2g$ & 3m & $p^4$ & $p_z$:&$A_1$ & $A_1@2g$&yes\\
        $(X)$ & $(00z)$ & & & $p_x,p_y$:&$E$ &  $E@2g$&yes \\
        \hline
        & $1e$& &&\multicolumn{3}{c}{\color{blue} $A_1'@1e$ : empty-site EBR}& \\
        & $(\frac{2}{3}\frac{1}{3}0)$  &  &  &  &  & \multicolumn{2}{r}{half-filled  }   \\
        \end{tabular}
        \end{ruledtabular}
\end{table}

\noindent{\bf RESULTS AND DISCUSSION} 
\paragraph*{Unconventional electronic band structure}
The 2H-$MX_2$ possesses a hexagonal structure with a space group (SG) of $P6_3/mmc$, where a hexagonal plane of $M$ atoms is sandwiched by two layers of $X$ atoms in Fig.~\ref{fig:1}(a). The $MX_2$ monolayers are connected by van der Waals force. For convenience, we mainly focus on the monolayer 2H-$MX_2$ (1H-phase). The $M$ and $X$ are located at the $1c$ and $2g$ Wyckoff positions of SG 187. Based on the atomic configurations, the atomic valence-electron band representations (ABRs) are generated by \webposabr in Table~\ref{table:ABRs}. 

The computed band structures of NbSe$_2$ and TaS$_2$ are presented in Figs.~\ref{fig:1}(b, c), respectively. The computed irreducible representations of energy bands indicate that the six lower energy bands belong to ABR $(A_1+E)@2g$, corresponding to the $p$ states of chalcogens in Table~\ref{table:ABRs}. An isolated band at $E_F$ belongs to  $A'_1@1e$ EBR with a half filling. The RSI of the empty $1e$ Wyckoff site is $\delta_1@1e\equiv m(A'_1)+m(A''_1)-m(A'_2)-m(A''_2)-m(E')+m(E'')=1$, with $m(\rho)$ is the number of EBR $\rho@1e$. Due to the presence of $C_3$ symmetry~\cite{unconphonon2023}, the Jenus 1H-TaSeS is also defined as an unconventional metal, and its superconductivity has been confirmed in experiment~\cite{yadav2023conventional}. \\

\paragraph*{Formation of the empty-site EBR at $E_F$} 
In a compound, all electronic states should originate from the ABRs~\cite{aBR2022}. The orbital-resolved band structures of representative NbSe$_2$ in Fig.~\ref{fig:2}(b) show the Fermi-level band is consist of Nb-$d_{z^2}$ and Nb-$d_{xy,x^2-y^2}$, which induce $A'_1@1c$ and $E'@1c$ ABRs, respectively. The band representations of topological quantum chemistry theory show 
\begin{equation}
A'_1@1c +E'@1c=A'_1@1e+E'@1e.
\end{equation}
Therefore, we conclude that the hybridization of the two ABRs gives rise to half occupied EBR $A'_1@1e$ and unoccupied EBR $E'@1e$, as illustrated in Fig.~\ref{fig:2}(a). In 1H-(Nb,Ta)$X_2$, the empty-site EBR $A'_1@1e$ is half filled, resulting in an unconventional metal. In the Na$_x$NbX$_2$ and Mo$_x$Nb$_{1-x}$X$_2$, this empty-site EBR becomes fully occupied at $x=1$, resulting in an unconventional insulator/obstructed atomic insulator~\cite{tqc2017,aBR2022,unconphonon2023,li2022obstructed,aBR2021}.  \\

\begin{figure}[!tb]
\centering
\includegraphics[width=8.5 cm]{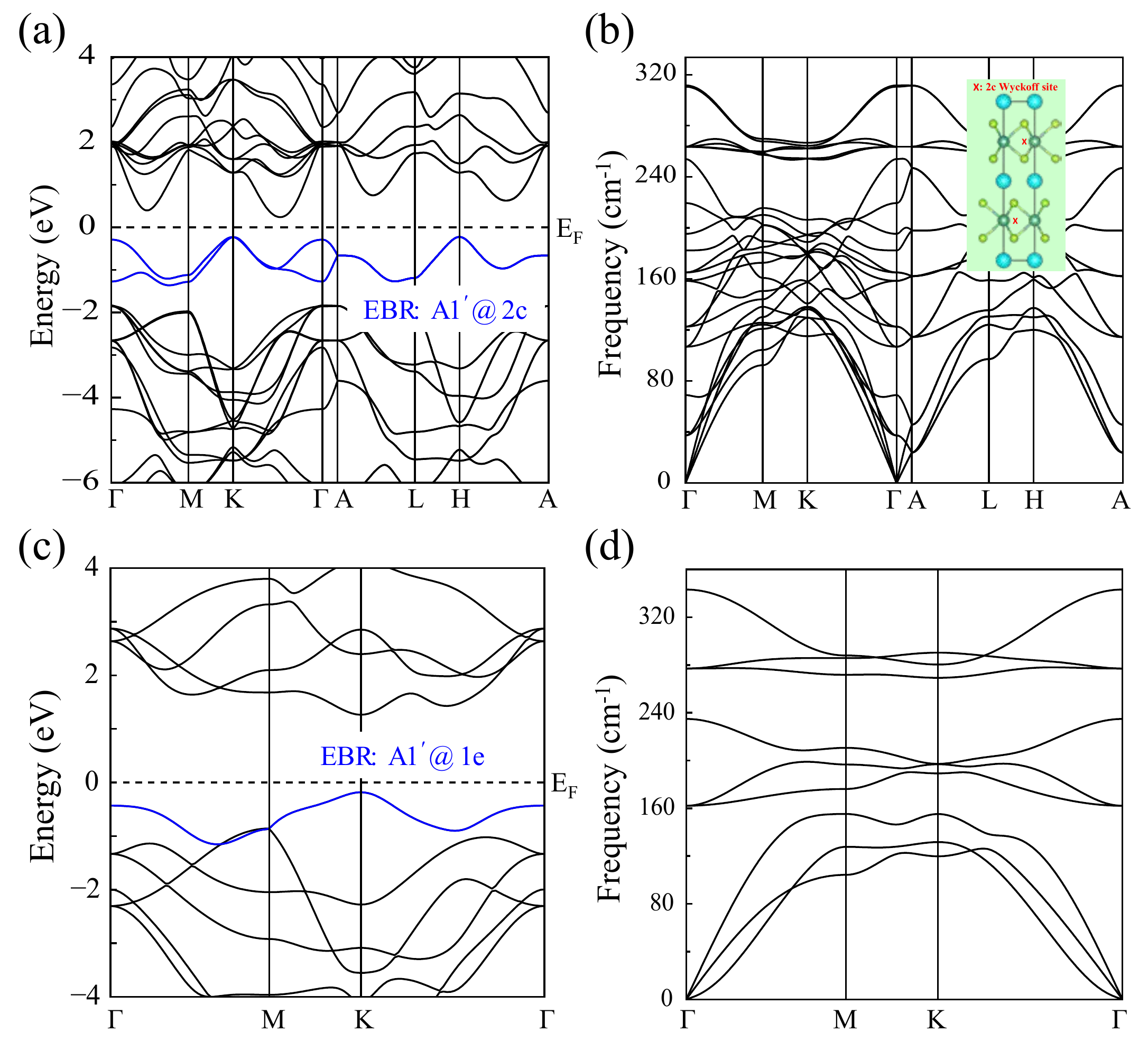}
\caption{Electronic structures and phonon dispersions after electron doping. Electronic structures and phonon dispersions of (a,~b) bulk NaNbSe$_2$ and (c,~d) 1H-MoSe$_2$. Inset in (b) presents the crystal structure of the bulk NaNbSe$_2$ with Na intercalation. The electron doping can stabilize the crystal structure.
} \label{fig:3}
\end{figure}

\paragraph*{CDW and electron doping}
We use different electronic smearing parameters to simulate different temperatures. The obtained phonon dispersions are presented in Figs.~\ref{fig:1}(d, e). At high temperatures, NbSe$_2$ and TaS$_2$ are stable with no negative frequency mode. While at low temperatures, there is a soft-mode band on $\Gamma$M, which is consistent with previous theoretical works~\cite{Weber2011,Calandra2009,lcs2022,Bianco2019}. This soft phonon mode is not caused by electron Fermi nesting. Instead, it is associated with EPC, and will lead to a CDW transition at low temperatures. To investigate electron doping effect, the electronic structures and phonon dispersions are computed for NaNb$X_2$ and 1H-Mo$X_2$. The crystal structure of NaNbSe$_2$ is shown in the inset of Fig.~\ref{fig:3}(b). It can be seen that Na atoms are intercalated between the layers. The $1e$ Wyckoff position in 1H-NbSe$_2$ corresponds to the $2c$ Wyckoff position in bulk NaNbSe$_2$. From its electronic structures and phonon dispersions in Figs.~\ref{fig:3}(a, b), it shows that the empty-site EBR $A'_1@2c$ is fully occupied and no negative phonon mode is found, suggesting a stable structure after Na intercalation. On the other hand, for Mo atoms substituting Nb atoms, the situation is similar: fully occupied empty-site EBR $A'_1@1e$ leads to the absence of the negative phonon mode, as shown in Figs.~\ref{fig:3}(c, d).
The results in the electron-doped compounds reveal that the presence of soft phonon mode is related to the half filling of the empty-site EBR. The electron doping can stabilize the crystal structure and suppress CDW transition. \\

\paragraph*{Superconductivity}
To investigate the SC property, we calculated the electron-phonon couplings $\lambda_{\bold{q}\nu}$ with $\sigma=0.02$ Ry, depicted by the magenta circles in Fig.~\ref{fig:2}(c). It shows that the soft phonon mode near $M$ point has the large $\lambda_{\bold{q}\nu}$. Further, the side and top views of the corresponding phonon vibration pattern at $M$ point are plotted in Fig.~\ref{fig:2}(d) to analyze the EPC-favorable vibration mode.  It shows that the soft phonon mode at $M$ point is mainly from Nb atoms that only vibrate in-plane.  On the right of Fig.~\ref{fig:2}(c), we calculated the Eliashberg spectral function $\alpha_2 F(\omega)$ and the  frequency-dependent coupling $\lambda(\omega)$. It can be seen that the contributions of 
\begin{equation} 
\lambda=\Sigma_{\textbf{q}v}\lambda_{\textbf{q}v}=2\int_{0}^{\infty}d\omega\frac{\alpha ^{2}F(\omega)}{\omega}  
\end{equation} 
mainly come from the phonon modes near 70 cm$^{-1}$. Due to the existence of soft phonon modes near $M$ point, the 1H-NbSe$_2$ hosts strong EPC ($\lambda > 1 $)~\cite{softphonon}. In the partially filling situation, the Fermi-level states are mainly centered at the empty site. On the other hand, this phonon mode strongly squeezes the empty site in Fig.~\ref{fig:2}(d). The coincidence between phonon mode and the Fermi-level states gives rise to the strong EPC, eventually leading to the SC instability. Besides, the SC in NbSe$_2$ can be tuned through carrier doping and dimensionality, such as Li/Na/Cu intercalation or Mo/W substitution~\cite{ni2htas2,li2htas2,org2hnbse2}.  \\

\paragraph*{Prediction of superconductivity in TaNS monolayer and TaN$_2$ bulk}

\begin{figure}[!t]
\centering
\includegraphics[width=8.5 cm]{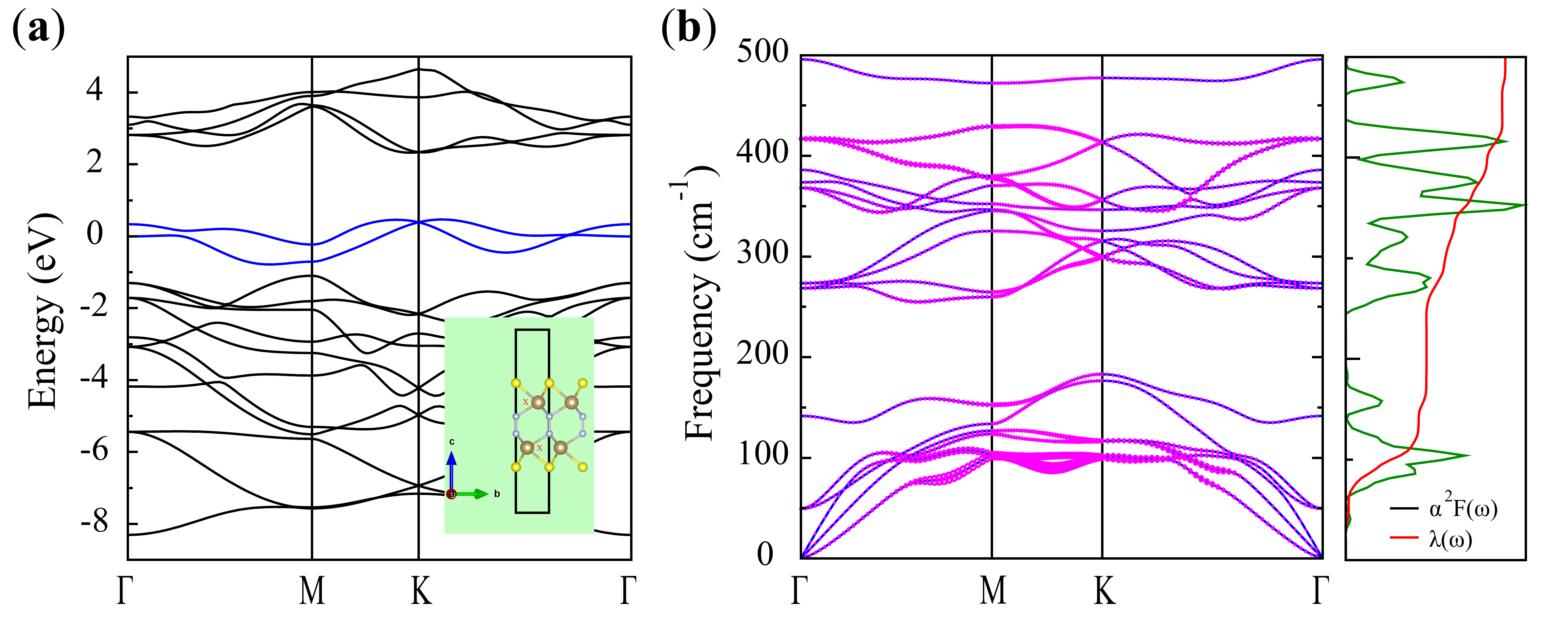}
\caption{Electronic structures, phonon dispersions, and electron-phonon couplings in TaNS monolayer. (a) Band structures of TaNS monolayer in SG 164. (b) Phonon spectrum, Eliashberg spectral functions $\alpha^{2}F(\omega)$, and the frequency-dependent coupling $\lambda(\omega)$. The electron-phonon couplings $\lambda_{\textbf{q}v}$ are represented by magenta circles.}\label{fig:4}
\end{figure}

As the partial filling of the empty-site EBR can lead the strong EPC and SC. Following this way, we predicted the SC in unconventional metallic TaNS monolayer [the inset of Fig.~\ref{fig:4}(a)]. 
The electronic structure is shown Fig.~\ref{fig:4}(a).  There exists a half-filling empty-site EBR at $E_F$, which is colored in blue. With the strong dimerization of N-N bonds, the band structure of TaNS shares a similar unconventional nature, with the electronic charge centers at empty sites (red crossings in the inset). The calculated phonon spectrum is shown in Fig.~\ref{fig:4}(b). There are no imaginary frequencies in the phonon spectrum, indicating the stability of the structure. From the calculated electron-phonon couplings $\lambda_{\bold{q}\nu}$, Eliashberg spectral functions $\alpha_2F(\omega)$, and the frequency-dependent coupling $\lambda(\omega)$, it can be seen that the EPC constants $\lambda$ are mainly contributed by the phonon modes near 100 cm$^{-1}$. The superconducting transition temperature ($T_C$) is estimated using Allen-Dynes modified McMillian equation~\cite{McMillan1968,Allen1975},
\begin{equation}\label{eq:Tc}
    \begin{split}
        T_C=\frac{\omega_{log}}{1.2k_{B}}\exp[\frac{-1.04(1+\lambda)}{\lambda(1-0.62\mu^{*})-\mu^{*}}]
    \end{split}
\end{equation}
where $k_{B}$ is the Boltzmann constant, $\mu^{*}$ is the effective screened Coulomb repulsion constant, typically $\sim$ 0.1, $\lambda$ is electron-phonon coupling constant, and $\omega_{log}$ is logarithmic average phonon frequency. With $\mu^{*}$ = 0.10 and $\lambda$ = 0.72, $T_C$ of TaNS is estimated to be 10 K using Allen-Dynes modified McMillian equation.

\begin{figure*}[!htb]
\centering
\includegraphics[width=17.5 cm]{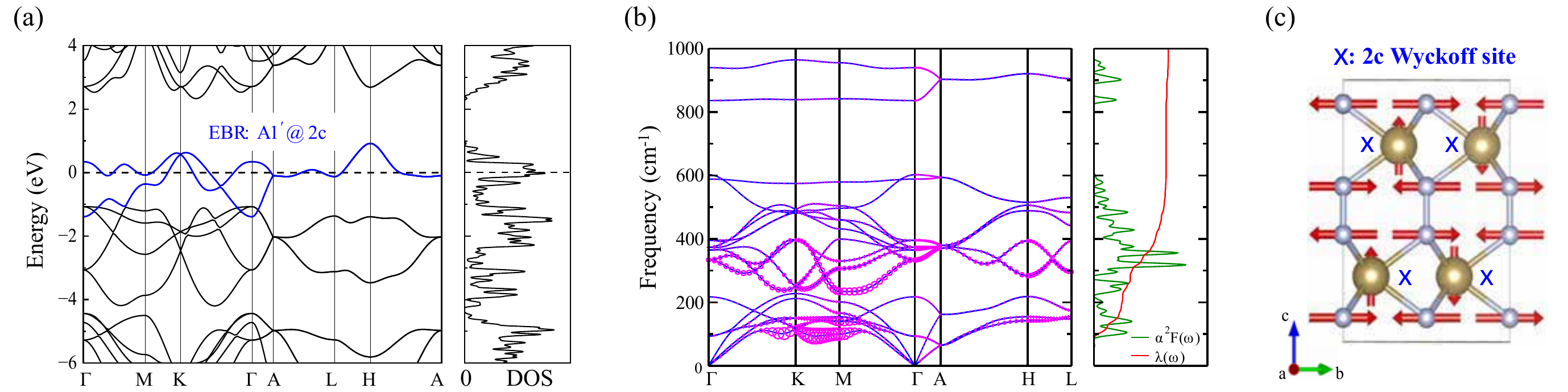}
\caption{Electronic structures, phonon dispersions, and electron-phonon couplings in bulk 2H-TaN$_2$. (a) Band structures and density of states of bulk 2H-TaN$_2$ in SG 194. The EBR of the low energy bands (in blue) is $A'_1@2c$, whose sites (indicated by ``x" in (c)) are empty in crystals. (b) Phonon spectrum, Eliashberg spectral functions $\alpha^{2}F(\omega)$, and the frequency-dependent coupling $\lambda(\omega)$. The electron-phonon couplings $\lambda_{\textbf{q}v}$ are represented by magenta circles. (c) The phonon vibration mode of lowest frequency at M.
}\label{fig:5}
\end{figure*}

Moreover, we predict another unconventional metal 2H-TaN$_2$ with superconductivity. Due to the dimerization of N-N bonds, they form the state of (N$_2$)$^{4-}$ in bulk TaN$_2$, resulting in the same situation of 2H-$MX_2$. The calculated  electronic structure for bulk 2H-TaN$_2$ is presented in Fig.~\ref{fig:5}(a). There exists a half-filling empty-site EBR, which is  colored in blue. 
The computed phonon spectra indicated that 2H-TaN$_2$ is stable with $\sigma=0.02$ Ry, as shown in Fig.~\ref{fig:5}(b). 2H-TaN$_2$ was also proposed to be a metastable phase at high pressure using structure prediction method~\cite{TaN2-HP}. With $\mu^{*}$ = 0.10 and $\lambda$ = 1.04, $T_C$ of bulk 2H-TaN$_2$ is estimated to be 26 K using Allen-Dynes modified McMillian equation, which is close to McMillan limit($\sim$ 40 K). As the $\lambda$ is compatible with conventional superconductors, the high $\omega_{log}$ is important in the final calculated $T_C$, which is mainly contributed from light mass N atoms. The phonon spectrum, and corresponding $\lambda_{\textbf{q}v}$, $\alpha ^{2}F(\omega)$, $\lambda(\omega)$ of 2H-TaN$_2$ are also shown in Fig.~\ref{fig:5}(b). The contributions of $\lambda(\omega)$ mainly come from the phonon modes of $\omega<$ 400 cm$^{-1}$. Among these phonon modes, we find that the low frequency phonon modes ($\sim$100 cm$^{-1}$) along $K$-$M$ high symmetry line show significant contribution to $\lambda$. The corresponding phonon vibration mode around $M$ is shown in Fig.~\ref{fig:5}(c). In this phonon mode, the Ta atoms vibrate along $z$ direction, and the N atoms vibrate in $xy$ plane, which has a higher amplitude than Ta atoms. It is noted that, the N atoms vibration modes along $z$ direction have frequencies higher than 500 cm$^{-1}$ because of strong N-N bonds. Thus, the in-plane vibration modes of N atoms couple with electride electron states around $E_F$, inducing high EPC constants.  \\

\paragraph*{Discussion}

The unconventional metals with an empty-site EBR encompass intriguing correlated states due to the soft phonon mode and strong EPC. We find that 1H/2H-phase TMD $MX_2$ has a half-filled empty-site EBR at $E_F$. The strong EPC is associated with half filling of the empty-site  $A'_1@1e$ EBR  near the Fermi level. The strong EPC may originate from the quantum geometric contributions of the unconventional electronic structure~\cite{quantumgeometry2023}; for example, the $\sigma$-bond band of MgB$_2$  belongs to the $A_{1g}@3f$ EBR at empty sites with partial filling. Our results indicate that the empty-site band in unconventional metals is crucial for electron-phonon coupling and superconductivity. Following this strategy, we predict unconventional metals TaNS monolayer and 2H-TaN$_2$ bulk  with SC $T_C=$ 10 K and  26 K respectively. It deserves further experimental focus and confirmation.  In conclusion, our findings reveal that the partial filling of the empty-site EBR in unconventional metals can give rise to strong EPC, as the Fermi-level states and phonon modes coincide spatially. The unconventional metals provide an ideal platform to search for superconductors. \\

\noindent{\bf METHODS} 
\paragraph*{Calculation method} 
We performed the first-principles calculations with QUANTUM ESPRESSO (QE) package~\cite{QE2009} based on the density functional theory (DFT) with the  projector-augmented wave (PAW) pseudopotentials~\cite{paw1,paw2}. The Perdew-Bruke-Ernzerhof (PBE) exchange-correlation functional of generalized gradient approximation was adopted. The dynamical matrices and electron-phonon coupling calculations were performed in the framework of density functional perturbation theory, as implemented in the QE package. The superconducting temperature was evaluated with Allen-Dynes modified McMillian equation using QE. The irreducible representations are computed by \webirpw~\cite{zhangsete}.  \\

\noindent{\bf DATA AVAILABILITY} \\
All data are available from the corresponding author on reasonable requests. \\

\noindent{\bf CODE AVAILABILITY} \\
All codes used for this work are open-source. The DFT calculations are based on QUANTUM ESPRESSO. IR2PW is available in \url{https://github.com/zjwang11/IR2PW}, which is used to compute the irreducible representation in QE. The POS2ABR is available at \url{https://github.com/zjwang11/UnconvMat}.  \\

\noindent{\bf ACKNOWLEDGEMENTS} \\
This work was supported by the National Key R\&D Program of China (Grants No. 2023YFA1607401, No. 2022YFA1403800), the National Natural Science Foundation of China (Grants No. 11974395, No. 12188101), the Strategic Priority Research Program of Chinese Academy of Sciences (Grant No. XDB33000000), and the Center for Materials Genome. \\

\noindent{\bf COMPETING INTERESTS} \\
The authors declare no competing interests. \\

\noindent{\bf AUTHOR CONTRIBUTIONS} \\
Z.W. proposed and supervised the project.
Z.Y., H.S., R.Z. and Z.G. conducted the theoretical calculations.  All authors contributed to analyzing the results and writing the manuscript.  \\

\noindent{\bf REFERENCES} \\

\bibliography{nbse}

\ \\

\end{document}